\documentclass[12pt]{article}
\usepackage{color,multicol}
\usepackage{amsmath}
\usepackage{amssymb}
\usepackage{epsfig}
\usepackage{epstopdf} 

\newcommand{\set}[1]{{\mathbb{#1}}}

\begin{document}

\title{Deriving Derivatives\footnote{Original version 23 April 2013. Journal-ref: Risk, July (2016), pp.\! 78-83.\vspace*{1mm}}}
\author{Andrei N.\ Soklakov\footnote{
Head of Strategic Development, Asia-Pacific Equities, Deutsche Bank.\vspace*{1mm}\newline
{\sl The views expressed herein should not be considered as investment advice or promotion. They represent personal research of the author and do not necessarily reflect the view of his employers, or their associates or affiliates.}
Andrei.Soklakov@(db.com, gmail.com).
}}
\date{}
\maketitle

\begin{center}
\parbox{14cm}{
{\small
Quantitative structuring is a rigorous framework for the design of financial products.
We show how it incorporates traditional investment ideas while supporting a more accurate expression of clients' views.
We touch upon adjacent topics regarding the safety of financial derivatives and the role of pricing models in product design.} }
\end{center}

\section{Introduction}

The ability to design high-quality products is the Holy Grail of modern (knowledge-based) industries. This ability is powered by a heavy use of science.

Scientific developments are regulated by the so-called {\it correspondence principle}, according to which all new theories must agree with their predecessors in those circumstances where the old approach remains accurate. As a famous inspirational example, we can mention Einstein's equations of General Relativity. These equations are numerically tuned to agree with Newton's laws in the limit of slow motion in weak gravitational fields. This knowledge-preserving compatibility is enforced despite the fact that the two theories work in completely different ways and even contradict each other regarding the most basic concepts such as space and time.

In quantitative structuring financial products are {\it derived} as solutions to clients' needs~\cite{Soklakov_2014_WQS}. This is very different from the historical approach where the precise structure of most financial products is best described as {\it ad hoc:} intuition-based  solutions, inspired, rather than determined, by clients' needs.

Although we argue for the greater use of quantitative methods, the historical approach remains enormously relevant. Indeed, it contains a huge amount of valuable experience. In order to preserve this experience we must follow the correspondence principle and connect quantitative structuring to the more familiar {\it ad hoc} approach. Of course, we must expect limits to how well the {\it ad hoc} legacy products can be justified, but, at the very least, we must be able to verify some intuitions behind their design.

In this paper we examine some of the iconic investment structures from the point of view of quantitative structuring.\\

The paper is organized as follows. After a short introduction to quantitative structuring, we walk through a series of examples. The examples are selected on pedagogical grounds: simple, yet with some general importance. We start with simple vanilla structures -- spot, volatility, skew products -- before gradually moving on to path-dependent exotics, index design and the feature of early exercise.

Our approach to structuring has powerful capabilities of product customization. This includes differentiating clients' views by strength and combining views on different risk factors. These customization capabilities, let us call them {\it view calculus}, generalize the presented set of individual illustrations. To broaden horizons even further we open a discussion on the connection between product design and modeling.

This paper invites but leaves unanswered two important questions. Firstly, we need to learn how to quantify investments in terms of their performance. Secondly, we need to show how the connection between quantitative structuring and modeling translates into concrete practical applications. We develop both of these topics in Ref.~\cite{Soklakov_2014MR} and give further illustration of the quantitative power of our approach in Ref.~\cite{Soklakov_2014EqPuzzle}.

\section{Quantitative Structuring -- the basics}

Quantitative structuring is a mathematical approach to product design. In this approach payoff structures are explicitly derived from the actual needs and views of the client.

At this moment, our framework is limited to the design of investment products. In terms of applications this is already a very large area.
It turns out, for example, that our framework can be used as a generic tool for risk assessment which employs financial concepts (such as payoff functions and investment performance) for understanding both the types and the materiality of risks~\cite{Soklakov_2014MR}.

Investment structuring is about incorporating knowledge into an investment structure. The Black-Litterman model is a good example of doing this in the portfolio framework of asset allocation~\cite{BlackLitterman_1992}. Quantitative structuring does the same in the realm of derivatives. More precisely, we are looking for payoff functions, $f(x)$, which define financial products as derivatives of some given underlying variables,~$x$. The variables can be as simple as the returns of individual stocks or as complex as the values of Markowitz or Black-Litterman portfolios.  The payoff functions are optimized around the client's needs and their views about the market.

In the fields of probability and statistics the acquisition of new knowledge is described using the well-known concept of a likelihood function. Naturally, the payoff function of an optimal investment turns out to be connected to the likelihood function describing investment research~\cite{Soklakov_2014_WQS}. There is one particular type of investor for whom this connection is especially simple. This is the growth-optimizing investor -- the investor which seeks a product with the greatest expected rate of return. For the growth-optimizing investor the optimal payoff function and the likelihood function simply coincide~\cite{Soklakov_2011}.

To formulate this mathematically, take a growth-optimizing investor with an interest in some uncertain quantity $x$. The market implies a distribution $m(x)$ for the quantity. This is the prior distribution. Having done some research, the investor develops his own view on $x$. This view is the posterior, or the investor-believed distribution, $b(x)$. The shape of the growth-optimal payoff can be computed, as the likelihood ratio~\cite{Soklakov_2011}
\begin{equation}\label{Eq:f}
f(x)=\frac{b(x)}{m(x)}\,.
\end{equation}

For the illustrations presented in this paper, it is sufficient to understand the above case of the growth-optimizing investor. The general investor, presented in~\cite{Soklakov_2013b}, enjoys much greater understanding and therefore much finer control of financial instruments. For example, the general investor can see whether additional features in a financial product can be understood as rational adjustments in risk preferences or whether gambling is taking place~\cite{Soklakov_2013b}. Such questions go beyond the capabilities of the historical approach and, as such, lie outside the scope of this paper.

\section{Examples}
\subsection{Spot and Volatility trades}
Consider an investor who has a view on either the expected value or the volatility of an index. For such cases we demand that our arguments should be more than just clear and simple, they should fit on the back of an envelope. The below discussions of Figs.~1.A and 1.B illustrate this point.

A view that the expected value of the index should be higher than suggested by the market corresponds to a believed distribution which is biased towards higher values of the index (as illustrated by Fig.~1.A). The growth-optimal payoff~(\ref{Eq:f}) can be easily sketched just by looking at the market-implied and investor-believed distributions. In the small enough near-ATM region this is easily recognizable as the profile of a forward contract.

Similarly, a belief that the market underestimates volatility corresponds to a believed distribution which is wider but not as tall as the market-implied distribution (see Fig.~1.B). Again, we can sketch the growth-optimal profile and see that it works pretty much the same way as the textbook vanilla combinations -- straddles and strangles. These classical vanilla combinations can be considered as crude approximations of the growth-optimal payoff.

\subsection{Skew trades}

\subsubsection{Analytical example}
Moving on to more complex examples, let us see how we could help investors with views on the skew. For a simple, analytically tractable illustration, recall the definition of a skew-normal distribution~\cite{OHagan_and_Leonhard_1976}.

Let $\phi()$ denote the probability density for the standard normal variable and $\Phi()$ denote the corresponding cumulative distribution function:
\begin{equation}
\phi(x)=\frac{1}{\sqrt{2\pi}}e^{-\frac{x^2}{2}}\,,\ \ \ \ \Phi(x)=\int_{-\infty}^x\phi(t)\,dt\,.
\end{equation}

\includegraphics[width=\textwidth]{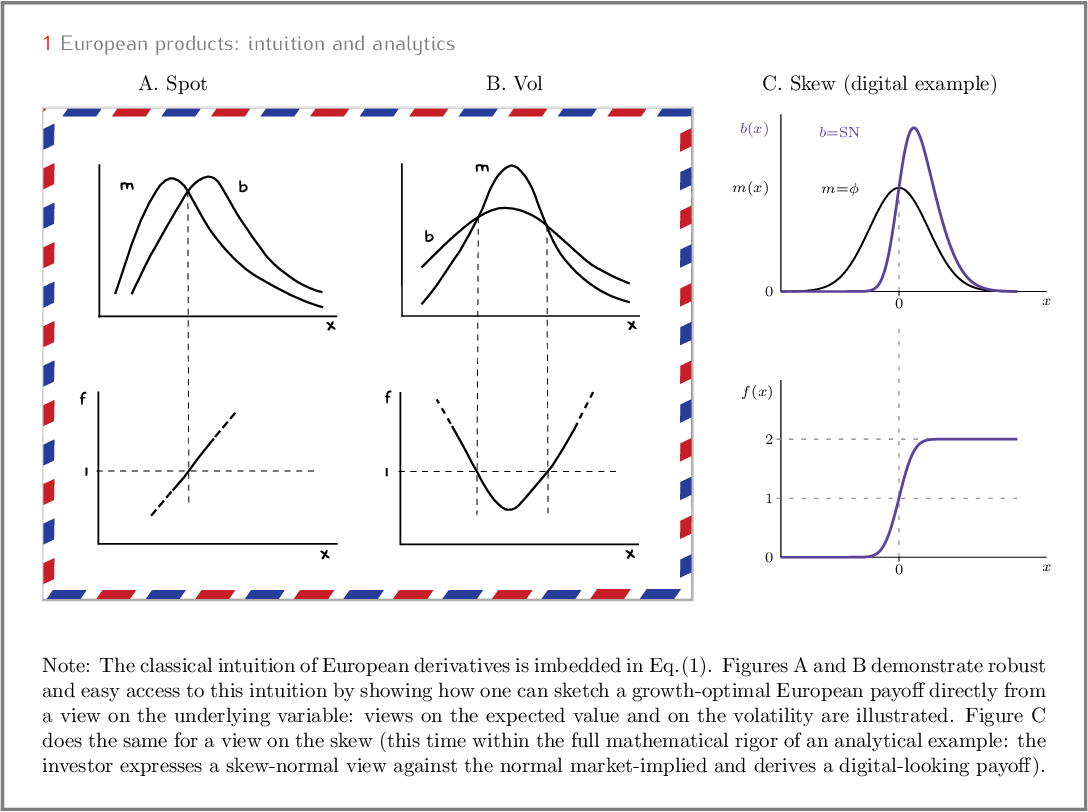}\\ \\

The probability density for a skew-normal random variable is defined as
\begin{equation}
{\rm SN}(x,\xi)=2\phi(x)\Phi(\xi x)\,,
\end{equation}
where $\xi$ is a parameter which controls the skew. The skewness (the third standardized moment) of this distribution is limited to the range between -1 and 1, so we should not expect great flexibility from the above analytical formulae. Nevertheless, this is a well-established example of introducing skew into the normal distribution, so we should know what it does in the context of our approach.

To this end, imagine a market which implies a normal distribution, $m(x)={\rm SN}(x,0)$, for some variable $x$. Now consider an investor which does not agree with the market. The investor believes that, in reality, there is a skew and uses a skew-normal distribution, $b(x)={\rm SN}(x,\xi)$, to describe this belief. For the growth-optimal payoff we compute
\begin{equation}
f(x)=\frac{b(x)}{m(x)}=\frac{{\rm SN}(x,\xi)}{{\rm SN}(x,0)}=2\Phi(\xi x)\,.
\end{equation}
We immediately recognize the profile of a cumulative distribution function (see Fig.~1.C) which contains a classic skew product -- the digital -- as a limiting case.

\includegraphics[width=\textwidth]{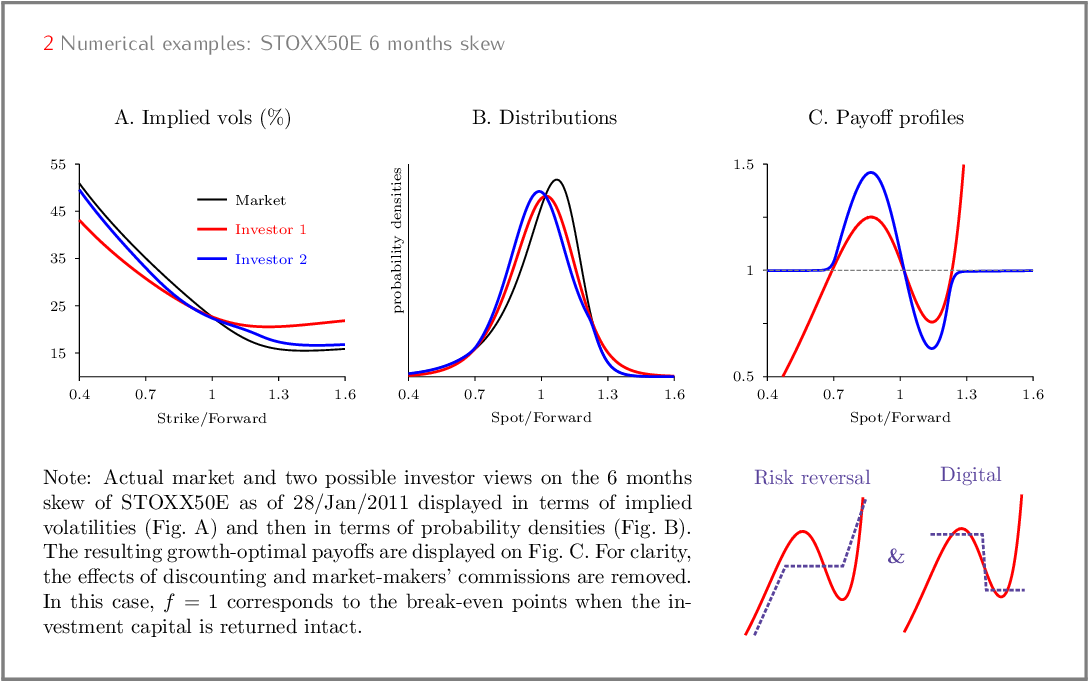}\\

\subsubsection{Numerical examples}\label{Sec:NumericalSkews}

Numerical examples based on real market data are just as easy to obtain. On Fig.~2.A we have the 6 months skew for STOXX50E as was implied by the market on 28/01/2011. Overlayed are two additional skew curves defining the beliefs of two possible investors. The investors practically agree on the skew slope near ATM. Both believe that it should be almost exactly half as steep as the market-implied.

The visible difference between the investors is in propagation of their near-ATM belief into the wings. Investor~1 takes a global view by tilting the entire volatility curve. Investor~2, by contrast, limits his view to the ATM region and does not challenge the market in the wings. Figure~2.B shows the relevant market and believed distributions. Finally, Figure~2.C displays the shapes of the growth-optimal payoffs as computed from Eq.~(\ref{Eq:f}).

We look at Investor~2 in greater detail later in the paper when we discuss product safety.

Here let us focus on the growth-optimal profile for Investor~1. We immediately recognize it as a smooth solution which blends two of the most popular ideas in skew trading -- risk reversals and digitals. The same picture shows that the skew has something to do with the third moment (recall the typical shape of a third-order polynomial). We see that our solution incorporates all of the relevant classic intuition while working with real data.

Note that all of the above payoffs can be easily delivered to the market via the standard replication technique as a static portfolio of vanilla options (see Eq.~(1) in Ref.~\cite{CarrMadan_2001}).

\subsection{Path-dependency}
The examples considered so far might give an impression that our framework is limited to vanilla combinations. The framework is, of course, a lot more general than that. This is because we have a lot of freedom in choosing the underlying variable, $x$.

We demonstrate this flexibility by giving a couple of additional examples. First, we continue within the theme of skew trading and consider an investor which requires more elaborate solutions than we considered so far. In the second example we learn how to build indices -- one of the most popular current directions in product innovation.

\subsubsection{Skew exotics}
In the above examples of skew trading we chose $x$ to be the future value of an index. This is not the only possible choice. Nothing stops us from considering, for instance, various third moment estimators and see what products we would get by using these path-dependent variables as definitions of $x$. The choice of $x$ is very important and it remains the job of a structurer to determine what view the client actually has.

For a specific and a rather typical example, let us consider a client which {\bf (\,i\,)} is an expert in a certain equity, {\bf (\,ii\,)} wants to express an opinion on the skew of that equity but {\bf (\,iii\,)} does not want to have significant delta exposure to the equity spot at any point during the life of the trade. It is clear that no product which can be represented by a combination of vanillas would satisfy the client. Such products would be rejected on the account of large equity delta. It follows that the only hope to meet the needs of the client lies with path-dependent products.

Mathematically, this means that we need to consider variables which depend on the entire sequence $S_{t:T}$ of equity fixings between the start, $t$, and the end, $T$, of the investment period. The particular choice of $x$ depends on further discussions with the client. For instance, the client might have a very detailed view on the dynamics of the equity. In this case, we might want to define $x=S_{t:T}$, deduce the distributions $m(S_{t:T})$, $b(S_{t:T})$ and derive the payoff as before. At first glance, this might appear as a very theoretical idea. In fact, this kind of approach can lead to remarkably simple and practical constructions. Below, in the section on index design, we demonstrate exactly this kind of approach. For now we imagine that our hypothetical client did not express any strong views on the underlying dynamics and left it to us to suggest a reasonable path-dependent variable, $x=x(S_{t:T})$.

In search of a new skew-sensitive variable we come back to the product which we designed for Investor 1 (the red profile on Fig.~2.C). Our new investor does not like delta exposure, so we decide to hedge it away. Assuming locally-lognormal dynamics for the underlying, $S_t$, the residual exposure of a delta-hedged European product can be approximated by the well-known expression (see e.g. Eq.~(9) in Ref.~\cite{CarrMadan_2001})
\begin{equation}\label{Eq:HedgedEuropean}
P\&L:=\sum_{i}\Gamma_\$(S_i,K)\Big(\sigma^2_i-K^2\Big)\Delta t_i\,,
\end{equation}
where $\Gamma_\$$ is proportional to the Black-Scholes dollar gamma of the product, $\sigma_i^2\Delta t_i$ is the actual realized variance for the $i$th hedging interval and $K$ is the volatility value used for Black-Scholes hedging.

Equation~(\ref{Eq:HedgedEuropean}) suggests a definition of a skew sensitive variable $x:=P\&L$. This definition should be cleaned up to the point that it can be specified on a termsheet. As we are not aiming to promote any particular product we are not going to do it here. There should be no doubt, however, that functions as nice as the ones plotted on Fig.~2.C can be approximated by profiles that are analytically tractable in simple models such as Black-Scholes (that is all we need to use Eq.~(\ref{Eq:HedgedEuropean}) in practice).

So far in this example all we have is an interesting definition of an underlying variable. This is still very far from having a finished derivative product. Nevertheless, let us see if the above logic can already lead us to any of the existing products.

To this end we imagine a structurer who has just derived the red-line profile on Fig.~2.C and decided to proceed with reducing delta by changing the underlying variable as outlined above. The structurer makes three additional simplifying assumptions. First, looking at the shape of the red profile on Fig.~2.C, he decides to use a cubic polynomial to approximate it. Second, he decides to replace Eq.~(\ref{Eq:HedgedEuropean}) with an analogous equation based on a locally-normal (Bachelier) dynamics and, third, he decides to overhedge and sets $K=0$.

With these assumptions the dollar-gamma becomes the ordinary second derivative of the payoff. Acting on a cubic profile, this leaves only two terms (linear and a constant), and the structurer derives
\begin{equation}\label{Eq:LongGammaShortVariance}
    x=\sum_{i}(\alpha S_i-\beta)\sigma^2_i\Delta t_i=\alpha\underbrace{\sum_{i} S_i\sigma^2_i\Delta t_i}_{\rm gamma\ swap}-\beta \underbrace{\sum_{i}\sigma^2_i\Delta t_i}_{\rm variance\ swap}\,,
\end{equation}
where, for the red profile on Fig.~2.C, both $\alpha$ and $\beta$ are positive. This is of course the classic long-gamma short-variance swap strategy~\cite{BouzoubaaOsseiran_2010} which is often noted as a sophisticated (reduced-delta) programme of achieving skew exposure.\footnote{In practice, the long-gamma short-variance swap strategy uses constant values $\alpha$ and $\beta$. This, of course, undermines its ability to achieve the delta-neutral exposure across the lifetime of the product. More accurate derivations either from Eq.~(\ref{Eq:HedgedEuropean}) or from more elaborate schemes~\cite{AhmadWilmott_2005} would be interesting to investigate.} Once again we see that a classical investment structure (long gamma short variance swap) can be seen within our framework as a result of understandable but rather crude decisions of the structurer with a lot of room for improvement.

The most important improvement would be to remember that choosing a good underlying variable, such as $x$ in Eq.~(\ref{Eq:LongGammaShortVariance}), is really just the first step of product design. It is very unlikely that a simple linear exposure to such a variable would coincide with an optimal product. The structurer should always look at the market-implied and investor-believed distributions of $x$ and then {\it derive} the payoff function $f(x)$.\footnote{It turns out that the derivation of a growth-optimal payoff can serve as a standard intermediate step towards the optimal product for a large class of investors~\cite{Soklakov_2013b}.} An example of such a derivation for a path-dependent underlying together with the relevant performance analysis can be found in~\cite{Soklakov_2014MR}.\\

\subsubsection{Index design}
Proprietary indices are one of the few areas of innovation which has seen substantial growth even in the face of strong negative sentiments from the crisis-ridden markets. Such indices continue to provide access to clever investment strategies in a relatively safe way. Investors can pull out completely or adjust participation in such strategies almost instantaneously. In this sense, indices can be viewed as the simplest kind of products with early exercise features.

In this section we show how investors can express their views on the {\it dynamics} of a market variable. We end up deriving a proprietary index which appears to contain both classic and modern ideas on trading strategies and is very open to almost any kind of adjustment or generalization.

Let $S_t$ be the total-return version of some equity (i.e. all dividends are kept within $S_t$). As a variable, consider the sequence, $x_{1:n}$, of daily returns:
\begin{equation}\label{Eq:IndexDesign_1st}
    x_i=\frac{\Delta S_i}{S_i}\,.
\end{equation}
We suppose that the market-implied dynamics for the equity can be written in a familiar form of a locally log-normal process
\begin{equation}\label{Eq:Market_x}
x_i=r_i\Delta t_i+\sigma_i\sqrt{\Delta t_i}\,\epsilon_i\,,\ \ \ \ \epsilon_i \sim {\cal N}(0,1)\,,
\end{equation}
where $r_i$ and $\sigma_i$ are the drift and the volatility. Assuming that $S_t$ is a tradable asset in it own right, the famous arbitrage arguments demand that $r_i$ is numerically close to the risk-free rate. Imagine now an investor which agrees with the market on everything except the drift, $r_i$. Such an investor would disagree with the above equation and would instead write
\begin{equation}\label{Eq:Investor_x}
x_i=\mu_i\Delta t_i+\sigma_i\sqrt{\Delta t_i}\,\epsilon_i\,,\ \ \ \ \epsilon_i \sim {\cal N}(0,1)\,,
\end{equation}
where $\mu_i$ is the investor's estimate of the drift. Equations (\ref{Eq:Market_x}) and (\ref{Eq:Investor_x}) can be rewritten as the market-implied and investor-believed probability distributions
\begin{equation}
m(x_{1:n})=\prod_{i=1}^n\frac{1}{\sigma_i\sqrt{\Delta t_i}}\,\phi\Big(\frac{x_i-r_i\,\Delta t_i}{\sigma_i\,\sqrt{\Delta t_i}}\Big)\,,\ \ \ \ b(x_{1:n})=\prod_{i=1}^n\frac{1}{\sigma_i\sqrt{\Delta t_i}}\,\phi\Big(\frac{x_i-\mu_i\,\Delta t_i}{\sigma_i\,\sqrt{\Delta t_i}}\Big)\,.
\end{equation}
For the growth-optimal payoff we compute
\begin{equation}
f_n=\frac{b(x_{1:n})}{m(x_{1:n})}=\prod_{i=1}^n \exp\bigg(\,\frac{(x_i-r_i\,\Delta t_i)^2-(x_i-\mu_i\,\Delta t_i)^2}{2\sigma_i^2\Delta t_i}\bigg)\,.
\end{equation}
Recalling the definition $x_i=\Delta S_i/S_i$, we obtain by direct calculation
\begin{equation}\label{Eq:DiscreteGirsanov}
\frac{f_i}{f_{i-1}}=\exp\bigg(\frac{\mu_i-r_i}{\sigma_i^2}\frac{\Delta S_i}{S_i}+\frac{r_i^2-\mu_i^2}{2\sigma_i^2}\Delta t_i\bigg)\,.
\end{equation}
In order to get a more practical formula for the index $f_i$ we note that, starting with equations (\ref{Eq:Market_x}) and (\ref{Eq:Investor_x}), all our calculations are done to the first order in $\Delta t_i$.
Expanding the above equation and keeping the terms to the first order in $\Delta t_i$ (remember to replace $(\Delta S_i/S_i)^2$ with $\sigma^2_i\Delta t_i$), we obtain
\begin{equation}
\frac{f_i-f_{i-1}}{f_{i-1}}\approx\frac{\mu_i-r_i}{\sigma_i^2}\Big(\frac{\Delta S_i}{S_i}-r_i\Delta t_i \Big)\,.
\end{equation}

\newpage
This leads us to a definition of an index $I_i$ with the following structure
\begin{equation}\label{Eq:TheIndex}
\frac{I_i-I_{i-1}}{I_{i-1}}:=\underbrace{\frac{\mu_i-r_i}{\sigma_i}}_{3}\underbrace{\frac{1}{\sigma_i}}_{2}\underbrace{\Big(\frac{\Delta S_i}{S_i}-r_i\Delta t_i \Big)}_{1}\,.
\end{equation}
This structure has three easily recognizable parts. Reading backwards from Eq.~(\ref{Eq:TheIndex}), we see that this is (1) an excess return index which is (2) vol-targeted on a (3) investor-expected Sharpe ratio. The importance of quantities such as Sharpe ratio for index design needs no explanation. As for vol-targeting and excess return -- these are two of the most popular techniques which are currently used in index designs.

On the theoretical side, we note that the general form of our result, namely the combination of quantities
\begin{equation}\label{Eq:IndexDesign_last}
\frac{{\rm growth\ rate}}{{\rm volatility}^2}\ \cdot\ {\rm risky\ return}
\end{equation}
is very similar to the leverage factor in Merton portfolio theory~\cite{Merton_1969}. Merton's original derivations are a beautiful example of mathematical prowess. Unfortunately, the methods of stochastic control theory require advanced mathematical training and, even then, it is difficult to imagine a practical business strategy that would need someone to come up with new tailored derivations on a regular basis.

By contrast, we arrived at a similar kind of investment advice using basic mathematics which can be easily tailored or generalized. We can consider investors with a view on volatility or both the volatility and the drift. We can also consider different distributions for returns (not necessarily normal). The calculations are so simple that virtually nothing can stop us from carrying them out in pretty much the same way as we did above.

The concentration of ideas that comes out of the above simple calculation is encouraging: derivation of key techniques of index design and building practical alternatives for Merton portfolio theory -- each of these topics is interesting in its own right.

\section{Early exercise and other connections to modeling}

In the case of indices, the feature of early exercise is trivial: index re-balancing times provide frequent exit opportunities for the investor. Understanding more complex products, such as American options, requires the concept of an exercise region. It is best defined by its operational meaning as an exercise trigger: as soon as the underlying variable wanders into the region, an exercise decision is recommended or enforced.

Before any investment product is purchased, its exercise region must be identified and understood by the investor. Indeed, the exercise region simply describes the circumstances in which the product is expected to deliver. It is hard to imagine how a rational investment can go ahead without understanding these circumstances.

Fortunately, the exercise region can often be read directly from the termsheet. Take for instance a European put option. The exercise region is the range of stock values at maturity between zero and the strike. One-touch options or, more generally, auto-callable products are examples of early exercise products which also state their exercise region on the termsheet: the region consists of barrier-observation times paired with the values of the underlying beyond the barriers.

Sometimes, the exercise region is hidden under the veil of optionality. American options are a classical example. The exercise region for an American option cannot be read directly from the termsheet. It is dynamically linked to the value of the option (the continuation value). Consequently, understanding the payoff of an American option requires the ability to value it.

Let us now see how this fits into quantitative structuring. Imagine a growth-optimizing investor using a model to understand American put options. The investor selects one particular contract with the strike $K_A$, computes its exercise region as a function of model parameters and deduces the model-implied distribution $b(S_\tau)$ for the stock upon exercise. The investor forms a view by believing the results of this investigation. Under what conditions would it make sense for the investor to express the view by buying the American put?

The answer is clear: the market must imply a distribution for $S_\tau$ which is different from $b(S_\tau)$ in a very particular way. Specifically, the market implied distribution $m(S_\tau)$, investor-believed $b(S_\tau)$ and the payoff function $f(S_\tau)=(K_A-S_\tau)^+$ must satisfy the familiar equation
\begin{equation}
b(S_\tau)=f(S_\tau)\,m(S_\tau)\,,
\end{equation}
where we remember that discounting factors are being ignored throughout this paper for simplicity.

Apart from having to use a model to understand nontrivial exercise conditions, this is identical to the earlier considered case of European exercise. And just as we found in the European case~\cite{Soklakov_2011}, American vanillas do not strike us as ideal investment vehicles. Perhaps, vanillas should be reserved for hedging purposes -- the job for which they were originally designed.

We are used to thinking of pricing models as tools for managing already existing products. The above discussion points to a deeper connection between modeling and product design. Below we give another example illustrating this connection from a very different angle. To this end we return back to our example of index design, Eqs.~(\ref{Eq:IndexDesign_1st}-\ref{Eq:IndexDesign_last}), and consider an alternative way of deriving the same index. In doing so we bring into our discussion some of the most common assumptions of quantitative finance.

Recall that our index investor had a view on the drift of a stochastic process. Note further that Eq.~(\ref{Eq:f}) can be interpreted as a change of measure: from market-implied to investor-believed. The growth-optimal payoff is the Radon-Nikodym derivative which can be computed using Girsanov theorem. This, of course, leads us to the same index. Indeed, the continuous versions of~Eq.(\ref{Eq:Market_x}-\ref{Eq:Investor_x}) can be understood as
\begin{eqnarray}
\frac{dS_t}{S_t}&=&r_t\,dt+\sigma_t\,dW_t^{\set{M}}\cr
                &=&r_t\,dt+\sigma_t\Big(\frac{\mu_t-r_t}{\sigma_t}\,dt + dW_t^{\set{B}}\Big)\,,
\end{eqnarray}
where $W_t^{\set{M}}$ and $W_t^{\set{B}}$ are Wiener processes in the market measure and in the believed measure respectively. By Girsanov theorem the payoff between the time $t$ and $t+\Delta t$ is
\begin{align}
f_t=\frac{d\,\set{B}}{d\,\set{M}}=&\exp\Big(\int_t^{t+\Delta t}\frac{\mu_u-r_u}{\sigma_u}\,dW_u^{\set{M}}-\frac{1}{2}\frac{(\mu_u-r_u)^2}{\sigma_u^2}\,du\Big)\cr
=&\exp\Big(\int_t^{t+\Delta t}\frac{\mu_u-r_u}{\sigma_u^2}\,\frac{dS_u}{S_u}+\frac{r_u^2 - \mu_u^2}{2\sigma_u^2}\,du\Big)\,,
\end{align}
which, of course, has the same meaning as the discrete~Eq.~(\ref{Eq:DiscreteGirsanov}).

Imagine now that we try to generalize the index for, say, volatility trading. An interesting observation emerges: the simple logic of Eqs.~(\ref{Eq:IndexDesign_1st}-\ref{Eq:IndexDesign_last}) turns out to be much more powerful than the measure-theoretic approach which we just introduced as an alternative. Indeed, a rewrite of Eqs.~(\ref{Eq:IndexDesign_1st}-\ref{Eq:IndexDesign_last}) on the account of different volatilities or even entirely different probability densities (not necessarily Gaussian) is a simple exercise in calculus.

By contrast, going via Girsanov theorem, we find that a change in drift is all we can do. Sophisticated measure theory gives us a beautiful continuum limit, but at a price. It locks us from understanding any investment product which cannot be explained as a view on drifts. Could this be adding to the pressure of using more risk factors (extra drifts) in modeling? Should we not prioritize more parsimonious discrete models? Before using a model for pricing a product, should we not check that the model is capable of capturing the trading rationale behind the product?

One thing is clear, quantitative structuring is giving us a new perspective on modeling. We exploit this fact in~\cite{Soklakov_2014MR} by developing a very general quantitative way of assessing model risk. In the same paper we learn how to assess materiality, i.e. how to measure performance of our products. We test the quantitative capabilities of our approach further by tackling one of the hardest problems in contemporary Economics -- the Equity Premium Puzzle~\cite{Soklakov_2014EqPuzzle}.

\section{View calculus and the origins of safety}

The above examples illustrate how classical intuitions of product design emerge within a more advanced framework of quantitative structuring in a variety of settings. In terms of progress, it is clear that the views of our clients can now be expressed much more accurately.

This ability of quantitative structuring to articulate clients' views (view calculus) overlaps with another very important topic concerning the safety of financial products. Below we summarize view calculus in terms of its basic capabilities and limitations. In particular, we note that product safety is a complex topic which cannot be covered using view calculus alone.

{\bf View differentiation} -- good financial products must reflect not just the general direction but also the strength of the clients' views. This is achieved by all of our examples. Even the back-of-an-envelope sketching technique which we used at the beginning of this paper is already powerful enough to see that. Very weak views lead to bond-like structures (no dependence on the underlying) and the stronger the view the more leveraged is the payoff.

\newpage
{\bf View integration} -- product structures must be able to combine all views of the client. Views on spot, volatility, skew or anything else must be integrated together. Our investor does not need to make any special arrangements to achieve that. This is done auto\-matically by thinking in terms of the believed distribution (which captures the net effect of all the views of the client).

{\bf View extrapolation} -- products should not extrapolate the actual views of the client. The client should get a product which expresses their exact view with no further exposures. Avoiding unnecessary risks has to be the most basic principle of any safety framework.

To see how this last point is achieved in our approach let us once again return to the skew trading example of section~\ref{Sec:NumericalSkews} and contrast Investors 1 and 2 on Fig.~2. Recall that Investor~2 has a tapered view outside the ATM region so as not to challenge the market in the wings. As a result, the client's exposure in the wings is capped.

Although the exposure capping effect is very robust, a word of caution is in order, for we came very close to the limits of intuition which Eq.~(\ref{Eq:f}) can provide. Note that the Investor 2 payoff has a comparatively higher amplitude near ATM. Investor~2 does not just curb his exposure in the wings, he reallocates the risk to the ATM region. Although not evident from Fig.~2.A, Investor~2 takes more risk near ATM than Investor~1.

To understand these effects and to learn how to control risk appetite we need to extend our approach to include techniques of risk aversion. We do this in Ref.~\cite{Soklakov_2013b}. Here we just mention that this goes far beyond the intuition that the historical approach (or even Eq.(\ref{Eq:f})) can support.
Investor~2 was built not by naive tapering of the views as Fig.~2.A might suggest (that would be surprisingly hard in practice). He was born from Investor~1 by using the techniques of Ref.~\cite{Soklakov_2013b}. In this particular case, the investor chose a near-ATM region (by fixing its boundaries near the outer break-even points of the payoff) and then revised his risk appetite: increasing risk tolerance within the region and expressing huge risk aversion outside the region. The revised payoff which followed from this change in risk appetite was then used to compute the equivalent view tapering of Fig.~2.A. The point is that none of this is evident from Fig.~2.A, so view tapering is not a good way of controlling risk appetite despite its intuitive appeal (see Ref.~\cite{Soklakov_2013b} for details).


\end{document}